\documentclass[aps,prd,preprintnumbers,showpacs,superscriptaddress,nofootinbib,amsmath,amssymb,floats,floatfix,showkeys,notitlepage,longbibliography]{revtex4-1}
\usepackage[table]{xcolor} 
\usepackage{array} 
\definecolor{lightblue}{rgb}{0.62, 0.83, 0.96} 
\definecolor{lightgray}{rgb}{0.93, 0.97, 0.99} 
\usepackage{comment}
\usepackage{graphicx}
\usepackage{subfigure}
\usepackage{palatino}
\usepackage[commandnameprefix=always]{changes}
\usepackage{hyperref}
\hypersetup{colorlinks=true,linkcolor=red,urlcolor=red,citecolor=red}
\usepackage[toc,page]{appendix}
\usepackage[normalem]{ulem}
\usepackage{lipsum}
\usepackage{graphicx}
\usepackage{subfigure}
\usepackage{palatino}
\usepackage{float}
\usepackage{sans}
\usepackage{adjustbox}
\usepackage{latexsym}
\usepackage{amsmath}
\usepackage{amssymb}
\usepackage{amsfonts}
\usepackage{dcolumn}
\usepackage{bm}
\usepackage{tikz}
\usepackage{bigints}
\usepackage{array,tabularx,multirow,booktabs}
\usepackage[tracking=true]{microtype}
\SetTracking{}{500}
\SetTracking{encoding={*}, shape=sc}{40}
\allowdisplaybreaks
\usepackage{adjustbox}
\usepackage{latexsym}
\usepackage{amsmath}
\usepackage{amssymb}
\usepackage{amsfonts}
\usepackage{dcolumn}
\usepackage{bm}
\usepackage{tikz}
\usepackage{bigints}
\usepackage{array,tabularx,multirow,booktabs}
\usepackage[tracking=true]{microtype}
\usepackage{color}
\allowdisplaybreaks

\begin{document}

\title{Reconciling the Weak Gravity and Weak Cosmic Censorship Conjectures in Einstein-Euler-Heisenberg-AdS Black Holes}

\author{Mohammad Reza Alipour}
\email{mr.alipour@stu.umz.ac.ir}
\affiliation{Department of Physics, Faculty of Basic
Sciences, University of Mazandaran\\ P. O. Box 47416-95447, Babolsar, Iran}
\affiliation{School of Physics, Damghan University, Damghan 3671645667, Iran}

\author{Saeed Noori Gashti}
\email{saeed.noorigashti@stu.umz.ac.ir}
\affiliation{School of Physics, Damghan University, Damghan 3671645667, Iran}

\author{Behnam Pourhassan}
\email{b.pourhassan@du.ac.ir}
\affiliation{School of Physics, Damghan University, Damghan 3671645667, Iran}
\affiliation{Center for Theoretical Physics, Khazar University, 41 Mehseti Street, Baku, AZ1096, Azerbaijan}
\affiliation{Centre of Research Impact and Outcome, Chitkara University, Rajpura-140401, Punjab, India.}

\author{\.{I}zzet Sakall{\i}
}
\email{izzet.sakallı@emu.edu.tr}
\affiliation{Physics Department, Eastern Mediterranean
University, Famagusta, 99628 North Cyprus, via Mersin 10, Turkiye}

\begin{abstract}
The potential conflict between the Weak Gravity Conjecture (WGC) and the Weak Cosmic Censorship Conjecture (WCCC) poses a significant challenge in general relativity. The WCCC serves as a fundamental assumption ensuring the coherence of gravitational theory. This study investigates the reconciliation of the WGC and the WCCC by examining Einstein-Euler-Heisenberg-AdS black holes in four-dimensional spacetime. By imposing specific constraints on the metric parameters, we demonstrate that the WGC and the WCCC can coexist harmoniously. Detailed analyses of Einstein-Euler-Heisenberg-AdS black holes for \( Q > M \) validate the simultaneous fulfillment of the two conjectures, particularly in scenarios where \( q^2/m^2 \geq \left(Q^2 / M^2\right)_{\text{e}} \). The electromagnetic self-interaction parameter \( \mu \) plays a crucial role in achieving this compatibility. Our findings establish that Einstein-Euler-Heisenberg-AdS black holes provide a robust framework for harmonizing the WGC and the WCCC. In particular, for exceedingly small values of ($\mu$)—or, equivalently, when the condition ($\mu \ll \ell$) is satisfied—the structure of our black hole transitions in a way that distinctly reveals its compatibility with the WGC. This study also explores the compatibility of the WGC and the WCCC with photon spheres. It examines parameter spaces that satisfy both conjectures, ensuring event horizons and photon spheres while maintaining black hole properties. Key results demonstrate that small \( \mu \) values preserve WCCC adherence and validate WGC through photon sphere characteristics.
\end{abstract}

\date{\today}

\keywords{Weak Gravity Conjecture; Weak Cosmic Censorship Conjecture; Einstein-Euler-Heisenberg-AdS black holes}

\pacs{}

\maketitle
\section{Introduction}\label{s1}
Quantum gravity stands as one of the most intriguing and formidable challenges in modern theoretical physics, captivating researchers with its vast implications for understanding the universe at its most fundamental level. The swampland program has emerged as a prominent research endeavor within this domain, aiming to establish universal principles that any consistent theory of quantum gravity must satisfy \cite{1,2,3,4,5,6}. Rooted in the premise that not all low-energy effective field theories are compatible with quantum gravity frameworks, such as string theory, the program seeks to distinguish viable theories within the string landscape from those relegated to the swampland. Drawing inspiration from areas like black hole physics, the AdS/CFT correspondence, and string theory constructions, the swampland program aspires to illuminate the profound nature of quantum gravity and its far-reaching impacts on cosmology and particle physics \cite{1,2,3,4,5,6,7,8,isz01,isz02}.

The AdS/CFT correspondence, a foundational duality connecting gravitational theories in anti-de Sitter (AdS) space with conformal field theories (CFT) on their boundaries, has proven to be a powerful instrument for probing the holographic nature of quantum gravity. This duality has provided a fertile testing ground for numerous swampland conjectures, including the WGC, the Distance Conjecture, and limits on the number of massless modes \cite{isz03,isz04}. Among the guiding principles of the swampland program is the absence of global symmetries in quantum gravity, with gauge symmetries as the sole exceptions. This requirement underpins the WGC, which posits the existence of particles with charge-to-mass ratios greater than one (\( q/m > 1 \)) in any quantum gravitational framework \cite{1,2,3,4,5,6,7,8}. This conjecture asserts that gravity is the weakest of all forces and serves as a cornerstone in identifying field theories consistent with quantum gravity. Researchers have explored the WGC's implications across diverse cosmological contexts, including black hole thermodynamics, black branes, and inflationary models \cite{a,b,c,d,e,f,g,h,i,j,k,l,m,n,o,p,q,r,s,t,u,v,w,x,y,z,aa,bb,cc,dd,ee,ff,gg,hh,ii,kk,ll,mm,nn,oo,qq,rr,ss,tt,uu,vv,ww,xx,yy,zz,aaa,bbb,ccc,ddd,isz05,isz06}. Equally significant is the WCCC, proposed by Penrose to preserve the causal structure and predictability of general relativity. The WCCC asserts that singularities arising from gravitational collapse are always concealed behind event horizons. However, achieving simultaneous compatibility between the WGC and WCCC has posed a longstanding challenge. For black holes, the WCCC is violated when \( Q > M \), exposing a naked singularity. Conversely, in the extremal case (\( Q = M \)), the WGC is satisfied due to decay products with \( q/m > 1 \), but the WCCC is violated as these products cannot form black holes \cite{9,10,isz07}. Recent investigations suggest that factors such as dark matter and a cosmological constant can act as stabilizing influences, preventing black holes from overcharging—a phenomenon unattainable in a vacuum. A critical threshold exists where the cosmological constant's effects become dominant, ensuring adherence to the WCCC in both linear and non-linear accretion scenarios. These findings underscore the pivotal role of cosmic factors in preserving the WCCC \cite{9,10,isz08}. Extending these insights, studies on a range of black holes, including extremal magnetized Kerr-Newman, dyonic Kerr-Newman, Einstein-Maxwell-Dilaton-Axion, Einstein-Born-Infeld, and rotating quantum BTZ black holes, have provided intriguing results. Nonetheless, the simultaneous compatibility of the WGC and WCCC remains elusive in many models, with conflicting compatibility ranges \cite{5000,1000,1001,1002,1003,1004,isz09,isz10}. It is also essential to note that while normal and subextremal black holes exhibit clear characteristics, the core framework and analysis of the WGC are founded on the assumptions involving extremal and superextremal states. This observation leads to the significant conclusion that the simultaneous validity of both the WGC and cosmic censorship hypothesis in most black hole models is challenging to achieve. This, in turn, necessitates a specialized classification of black holes to address these complexities.

We propose classifying black holes from the perspective of the WGC while earlier ensuring adherence to the laws of thermodynamics, addressing contradictions, and thoroughly comparing available observable data. Black holes embedded within the framework of string theory—such as Gauss-Bonnet black holes, Brane black holes, and charged black holes with additional structures like quintessence, string clouds, or perfect fluids—can be systematically categorized in this way. Also, extending this classification to other models, such as PFDM black holes and similar structures, offers an intriguing avenue for analysis. This novel classification creates opportunities to investigate further evidence and strengthen the validation of the WGC. Indeed, charged black holes (e.g., those with quintessence, perfect fluid, or string cloud structures) in extremal or superextremal states can emit superextremal particles, such as electrons with charge-to-mass ratios exceeding one. This emission mechanism aligns with the WGC by ensuring that gravity remains the weakest force in these systems. By facilitating this decay process, the WGC effectively safeguards the stability and fundamental consistency of black holes \cite{isz10}. In other related studies, we have addressed this contradiction by examining black hole thermodynamics under different conditions, including extended phase space, restricted phase space, and conformal field theory (CFT) frameworks. These efforts have provided valuable insights and have been documented in recent publications \cite{vv}. Nonetheless, the tension between the WGC and cosmic censorship remains a critical challenge within the Swampland Program. Resolving this issue requires identifying and categorizing black hole models that demonstrate the highest degree of compatibility with both conjectures. This article aims to contribute to this endeavor by offering a systematic approach to black hole classification, informed by the interplay of the WGC, observable phenomena, and theoretical principles.\\

The main goal is to select black holes that satisfy many conditions and can be classified into our new classification from the perspective of the swampland program. This involves maintaining various conditions, solving a series of contradictions, and creating a new avenue for communication between the universe and quantum mechanics. This work can be expanded, and the model mentioned has the basic conditions in the initial investigations for other studies, such as the investigation of photon spheres and other cosmic concepts, while maintaining the WGC. This will help find more evidence for the swamp program and provide a new path for study, raising many questions that can be leveraged to advance the science of physics.\\

The remainder of this paper is organized as follows. In Sec.~\ref{sec2}, we introduce the Einstein-Euler-Heisenberg-AdS black hole model and derive its fundamental properties. Section~\ref{sec3} explores the connection between the Weak Gravity Conjecture and the Weak Cosmic Censorship Conjecture, establishing the conditions under which they can simultaneously hold. In Sec.~\ref{sec4}, we investigate the compatibility of the WGC with photon spheres, analyzing their topological properties and stability in relation to the black hole parameters. Finally, Sec.~\ref{sec5} presents our conclusions and discusses implications for future research in quantum gravity.

\section{The Model}\label{sec2}
Within the framework of Euler-Heisenberg theory, the Lagrangian was constructed using Lorentz and gauge invariants, as follows \cite{500,504,505,506,507},
\begin{equation}\label{eq1}
\mathcal{L}_{\text{EH}} = -\frac{1}{4}F_{\mu\nu}F^{\mu\nu} + \frac{\mu}{4}\left[(F_{\mu\nu}F^{\mu\nu})^2 + \frac{7}{4}(-^*F^{\mu\nu}F_{\mu\nu})^2\right],
\end{equation}
where \( \mu \) is the parameter representing electromagnetic self-interactions, given by,
\begin{equation}\label{eq2}
\mu = \frac{2\alpha^2}{45m_e^4},
\end{equation}
with \( \alpha \) being the fine structure constant and \( m_e \) the electron mass. Here, \( ^*F^{\mu\nu} = \frac{1}{2\sqrt{-g}} \epsilon_{\mu\nu\rho\sigma}F^{\rho\sigma} \) denotes the dual of the electromagnetic field tensor \( F_{\mu\nu} \) \cite{500,504,505,506,507}. The \( \mu \)-term in the Lagrangian introduces a nonlinear contribution to the gauge field equations via the squared electromagnetic field tensor \cite{500,504,505,506,507}.
The four-dimensional action of general relativity, coupled to nonlinear electrodynamics, is expressed as \cite{500,504,505,506},
\begin{equation}\label{eq3}
S_{\text{EH}} = \frac{1}{4\pi} \int_{M^4} d^4x \sqrt{-g} \left( \frac{R}{4} + \Lambda + \mathcal{L}_{\text{EH}} \right),
\end{equation}
where \( g \) represents the determinant of the metric tensor, \( R \) is the Ricci scalar, and \( \Lambda \) is the cosmological constant. This action gives rise to the Einstein-Euler-Heisenberg field equations. The solution to these field equations, in the case of a static spherically symmetric metric for an Euler-Heisenberg-AdS black hole in a strong electromagnetic field, is given by \cite{506,507,508,509},
\begin{equation}\label{eq4}
ds^2 = f(r) dt^2 - \frac{dr^2}{f(r)} - r^2 d\theta^2 - r^2\sin^2\theta d\phi^2,
\end{equation}
with the metric function,
\begin{equation}\label{eq5}
f(r) = 1 - \frac{2M}{r} + \frac{Q^2}{r^2} - \frac{\mu Q^4}{20r^6} + \frac{r^2}{\ell^2}.
\end{equation}
Here, \( M \) is the black hole’s mass, \( Q \) its electric charge, and \( l \) is the AdS radius related to the cosmological constant by \( \Lambda = -\frac{3}{l^2} \). The parameter \( \mu \) is the Euler-Heisenberg self-interaction constant \cite{506,508,509}. When \( \mu = 0 \), the solution reduces to the familiar Reissner-Nordström metric \cite{506,508,509}.

The nonlinear effects, primarily due to the \( \mu \)-term in the metric, arise from the fourth power of the electric charge and the inverse sixth power of the radial coordinate. These effects, originating from vacuum polarization, make the compact object more gravitationally attractive compared to the Reissner-Nordström-AdS black hole \cite{506,508,509}. Figures illustrating the behavior of the metric functions of Einstein-Euler-Heisenberg-AdS black holes in comparison with Reissner-Nordström black holes provide further insight into these dynamics.

Based on the findings in \cite{510}, the curves beyond the outer horizons of two types of black holes—those with and without electromagnetic self-interaction corrections—exhibit remarkable similarity. As a result, the motion of test particles or light around these gravitational sources may not clearly distinguish the subtle differences introduced by the electromagnetic self-interactions. This observation motivates a deeper comparison of the thermodynamic properties of Einstein-Euler-Heisenberg-AdS black holes with those of their counterparts that lack nonlinear effects.

An intriguing question arises: Could the unique internal structures of black holes modified by nonlinear effects influence their evolutionary behavior? Addressing this requires advancing research into the thermodynamics of black holes within the framework of nonlinear electrodynamics.

The horizon radii of such black holes can be determined as the real roots of the equation \( f(r) = 0 \). For the outer horizon, \( r_+ \), this condition can be explicitly expressed as \cite{510},
$f(r_+) = 0$. By combining this condition with Eq. (\ref{eq5}), the mass of the black hole is given as \cite{510,511}
\begin{equation}\label{eq6}
M = \frac{r_+}{2} \left( 1 + \frac{Q^2}{r_+^2} - \frac{\mu Q^4}{20r_+^6} + \frac{r_+^2}{\ell^2} \right),
\end{equation}
where \( Q \) represents the black hole's electric charge, \( \mu \) is the Euler-Heisenberg parameter capturing nonlinear effects, and \( l \) is the AdS radius associated with the cosmological constant (\( \Lambda = -\frac{3}{l^2} \)).
The Hawking temperature for this class of black holes can be derived as \cite{511}
\begin{equation}\label{eq7}
T = \frac{f'(r)}{4\pi} \Big|_{r = r_+} = \frac{1}{4\pi r_+} \left( 1 - \frac{Q^2}{r_+^2} + \frac{\mu Q^4}{4r_+^6} + \frac{3r_+^2}{\ell^2} \right).
\end{equation}
The entropy of the black hole is defined as \cite{512}
\begin{equation}\label{eq8}
S = \pi r_+^2,
\end{equation}
which can also be expressed in terms of the horizon’s area \( A = 4\pi r_+^2 \) using the standard relation \( S = \frac{A}{4} \) \cite{512,30'',31''}. These findings underscore the importance of further investigating the thermodynamic implications of nonlinear effects in black holes, as well as exploring how these effects influence their evolutionary dynamics and unique internal structures \cite{500}.

\section{WGC-WCCC connection}\label{sec3}
\begin{figure}[h!]
 \begin{center}
 \subfigure[]{
 \includegraphics[height=6cm,width=7cm]{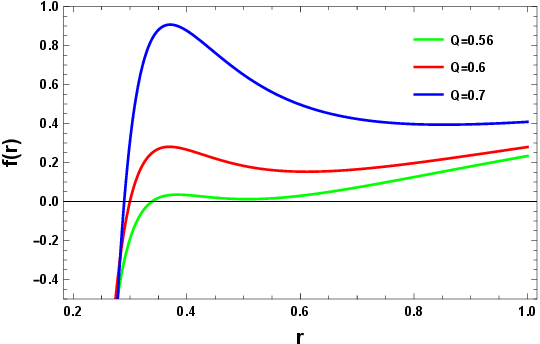}
 \label{fig1a}}
 \subfigure[]{
 \includegraphics[height=6cm,width=7cm]{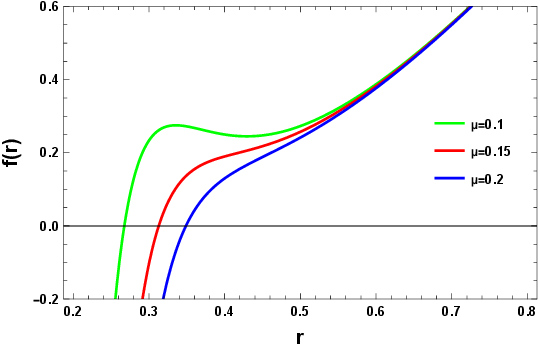}
 \label{fig1b}}
 \caption{\small{The metric function. (a) With the $\ell=7$, $M=0.55$ and $\mu=0.15$. (b) With the $\ell=1$, $M=0.55$ and $Q=0.56$.}}
 \label{m1}
 \end{center}
 \end{figure}

\begin{figure}[h!]
 \begin{center}
 \includegraphics[height=6cm,width=7cm]{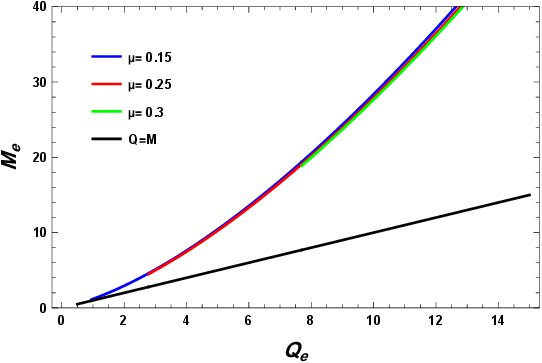}
 \caption{\small{The ($Q_{e}-M_{e}$) plan for different values of $\mu$ and $\ell=1$.}}
 \label{m2}
 \end{center}
 \end{figure}

\begin{figure}[h!]
 \begin{center}
 \subfigure[]{
 \includegraphics[height=5cm,width=6cm]{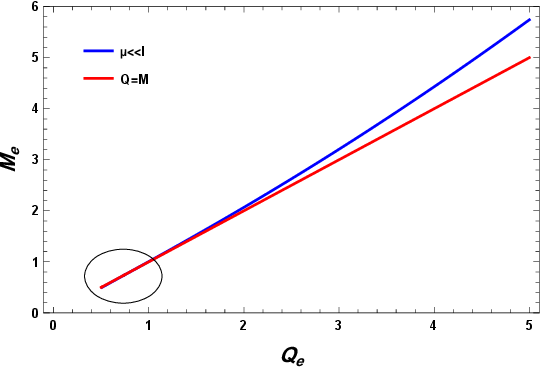}
 \label{fig2a}}
 \subfigure[]{
 \includegraphics[height=5cm,width=6cm]{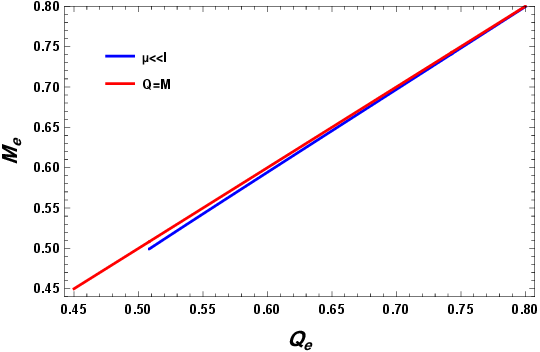}
 \label{fig2b}}
 \caption{\small{The ($Q_{e}-M_{e}$) plan. fig (a) With the $\ell=7$,  and $\mu=0.15$. (b): The circular part enlarged in fig (a)}}
 \label{m3}
 \end{center}
 \end{figure}

The WGC is an integral aspect of the "Swampland Program," a framework that distinguishes effective field theories (EFTs) compatible with quantum gravity from those that are not. This program categorizes EFTs into the "landscape," representing consistent theories, and the "swampland," which includes inconsistent ones. Within this context, the WGC acts as a guiding principle, ensuring that gravity remains the weakest force in $U(1)$ gauge theories by necessitating the existence of particles with a charge-to-mass ratio greater than one $(q/m > 1)$.
The implications of the WGC are particularly significant in the context of black holes. Charged black holes are classified into subextremal $(Q < M)$, extremal $(Q = M)$, and superextremal $(Q > M)$ categories. The WGC suggests that extremal black holes decay into superextremal particles, preventing the formation of naked singularities and maintaining consistency with the cosmic censorship hypothesis. Conversely, if the WGC is violated, black holes could transition into superextremal states, forming naked singularities that defy established physical principles.
The existence of superextremal particles is crucial for the stability and evaporation of black holes, highlighting their role in upholding the delicate balance of forces dictated by the WGC. Thermodynamic laws provide a valuable framework for analyzing black hole behavior, offering insights into their stability, decay processes, and overall compatibility with the conjecture.
To test the WGC, researchers investigate cosmic phenomena, contradictions, and black hole classifications, seeking observable evidence that aligns with theoretical predictions. This approach not only strengthens the foundation of the WGC but also bridges the gap between quantum mechanics and cosmology. Through this comprehensive investigation, the WGC emerges as a pivotal tool for understanding the universe's fundamental principles, deepening our grasp of the interplay between gravity, quantum theories, and cosmic behavior.
We study the metric of a Einstein-Euler-Heisenberg-AdS black holes to investigate its horizon structure and physical properties. The event horizon radius is found by solving the equation \( f(r) = 0 \), where \( M \) denotes the black hole's mass and \( Q \) its charge. For \( Q > M \), the black hole has no event horizon, leaving its singularity exposed to external observers. This situation, known as a naked singularity, contradicts the WCCC, which asserts that singularities must always be hidden by event horizons.
By zeroing the metric equation alongside the extremality condition, which is determined by zeroing its derivative or the zeroing temperature, we derive the conditions under which the two conjectures, namely the Weak Gravity Conjecture (WGC) and the Weak Cosmic Censorship Conjecture (WCCC), are satisfied. To account for the effects of the parameter \( \mu \), and above explanation, we can determine \( r_{\text{e}} \) using the following equations,
\begin{equation}\label{eq10}
\frac{r_e^2}{\ell^2}-\frac{2 M_e}{r_e}+\frac{Q_e^2}{r_e^2}+1=\frac{Q^4 \mu }{20 r_e^6},
\end{equation}
and,
\begin{equation}\label{eq11}
\frac{3 r_e}{\ell^2}-\frac{Q_e^2}{r_e^3}+\frac{1}{r_e}=-\frac{Q_e^4 \mu }{4 r_e^7}.
\end{equation}
Since the WCCC and the WGC are not universally valid for all black holes, we focus on a specific case where these conjectures cannot simultaneously hold without the parameter \( \mu \). By isolating the \( \mu \)-dependent terms in the metric, we aim to identify points where the WCCC can still be satisfied with WGC, at least at one extremal configuration.
Using a well-established relationship for the WGC, we calculate the extremal limit and determine the WGC range. We then employ numerical analysis to explore the compatibility range of the two conjectures. If a region exists where both conjectures are satisfied at critical or extremal points, this range may be extended to other regions of the black hole's parameter space.
By examining different types of black holes and categorizing them, we can identify candidates suitable for investigating quantum gravity concepts. These classified black holes provide a foundation for extensive research while preserving essential principles of physics, allowing a comprehensive examination of the swampland program.
We calculate \( r_{\text{e}} \) (as shown in Appendix A) using the derived equations, put it into the mass equation to determine the extremality limit, and establish the inequality \( q^2/m^2 > (Q^2/M^2)_{\text{e}} \). Due to the high-order terms involved in these equations, an analytical solution is impractical; hence, numerical methods are employed to identify compatibility points between the WCCC and WGC. The results, summarized in Tables (\ref{P1}) and (\ref{P2}), demonstrate that for very small values of \( \mu \)  ($\mu \ll \ell$), the two conjectures align. However, when \( \mu \) approaches zero or the upper limit, this compatibility is lost.
Fig. (\ref{m1}) illustrates the metric functions for various free parameters and different values of mass and charge. This serves as a foundation for examining the compatibility of the two conjectures in subsequent calculations.
As depicted in (\ref{m2}), the variations in ($Q_{e}-M_{e}$) are effectively visualized through the plotted diagrams. These illustrations aim to identify the compatibility range where the WGC holds true. As the parameter ($\mu$) approaches ($\ell$) and even exceeds it, the inconsistency between the two conjectures becomes more pronounced. Specifically, while the Weak Cosmic Censorship Conjecture (WCCC) remains valid in this scenario, the Weak Gravity Conjecture (WGC) exhibits noticeable inconsistency. This deviation can be attributed to the high values of $\mu$ or cases where $\mu$ is close to ($\ell$). Consequently, under such conditions, the two conjectures are not simultaneously satisfied, highlighting the sensitive dependence of their interplay on $\mu$'s magnitude.
However, as will be demonstrated in subsequent sections, this apparent inconsistency can be resolved by considering scenarios in which $\mu$ assumes smaller values. By analyzing such cases, we can establish conditions under which both WGC and WCCC are reconciled, thus restoring the consistency between these fundamental conjectures. This refined analysis underscores the critical influence of $\mu$ and offers deeper insights into the intricate dynamics governing their relationship. Specifically, for significantly small values of ($\mu$)—or equivalently, under the condition ($\mu \ll \ell$)—our black hole structure evolves in a manner that prominently exhibits compatibility with the WGC, as demonstrated in Fig. (\ref{m3}). Notably, Fig. (\ref{fig2b}) provides a magnified view of the circular region highlighted in (\ref{fig2a}), allowing for a clearer observation of this compatibility. As previously discussed, several compatible points can be determined via numerical analysis, and these findings are presented comprehensively in the two tables (\ref{P1} and \ref{P2}).

\begin{table}[h!]
\centering
\setlength{\arrayrulewidth}{0.4mm} 
\setlength{\tabcolsep}{2.8pt} 
\arrayrulecolor[HTML]{000000} 
\begin{tabular}{|>{\centering\arraybackslash}m{2cm}|>{\centering\arraybackslash}m{2cm}|>{\centering\arraybackslash}m{2cm}|>{\centering\arraybackslash}m{2cm}|>{\centering\arraybackslash}m{2cm}|>{\centering\arraybackslash}m{2cm}|>{\centering\arraybackslash}m{2cm}|>{\centering\arraybackslash}m{2cm}|}
\hline
\rowcolor[HTML]{9FC5E8} 
\textbf{$\ell$} & \textbf{$\mu$} & \textbf{$Q_e$} & \textbf{$r_{e}$} & \textbf{$M_{e}$} & \textbf{$(Q/M)_{e}>1$} & \textbf{$WGC-WCCC$}\\ \hline
\rowcolor[HTML]{EAF4FC} 
1 & 0.003 & 0.08 & 0.071908 & 0.07904 & 1.0121 & $\checkmark$ \\ \hline
1 & 0.003 & 0.09 & 0.08325 & 0.08933 & 1.00748 & $\checkmark$ \\ \hline
\rowcolor[HTML]{EAF4FC}
1 & 0.003 & 0.1 & 0.093803 & 0.099584 & 1.00417054 & $\checkmark$ \\ \hline
1 & 0.003 & 0.15 & 0.14262 & 0.150997 & 0.99339 & $\times$ \\ \hline
\rowcolor[HTML]{EAF4FC}
1 & 0.003 & 0.2 & 0.18788 & 0.2031943 & 0.9842 & $\times$ \\ \hline
1 & 0.003 & 0.5 & 0.407201 & 0.54391 & 0.919261 & $\times$ \\ \hline
\end{tabular}
\caption{The condition for consistency WGC-WCCC for $\mu=0.003$}
\label{P1}
\end{table}

\begin{table}[h!]
\centering
\setlength{\arrayrulewidth}{0.4mm} 
\setlength{\tabcolsep}{2.8pt} 
\arrayrulecolor[HTML]{000000} 
\begin{tabular}{|>{\centering\arraybackslash}m{2cm}|>{\centering\arraybackslash}m{2cm}|>{\centering\arraybackslash}m{2cm}|>{\centering\arraybackslash}m{2cm}|>{\centering\arraybackslash}m{2cm}|>{\centering\arraybackslash}m{2cm}|>{\centering\arraybackslash}m{2cm}|>{\centering\arraybackslash}m{2cm}|}
\hline
\rowcolor[HTML]{9FC5E8} 
\textbf{$\ell$} & \textbf{$\mu$} & \textbf{$Q$} & \textbf{$r_{e}$} & \textbf{$M_{e}$} & \textbf{$(Q/M)_{e}>1$} & \textbf{$WGC-WCCC$}\\ \hline
\rowcolor[HTML]{EAF4FC} 
1 & 0.1 & 0.61 & 0.400392 & 0.66332 & 0.919614 & $\times$ \\ \hline
1 & 0.1 & 0.7 & 0.4697355 & 0.782015 & 0.89512275 & $\times$ \\ \hline
\rowcolor[HTML]{EAF4FC}
1 & 0.1 & 1 & 0.62103289 & 1.208324 & 0.827591 & $\times$ \\ \hline
1 & 0.1 & 2 & 0.965552 & 2.95655 & 0.676463 & $\times$ \\ \hline
\rowcolor[HTML]{EAF4FC}
1 & 0.1 & 5 & 1.60842 & 10.51116 & 0.475685 & $\times$ \\ \hline
1 & 0.1 & 10 & 2.3133552 & 28.58306 & 0.34985 & $\times$ \\ \hline
\end{tabular}
\caption{The condition for consistency WGC-WCCC for $\mu=0.1$}
\label{P2}
\end{table}
\section{WGC with photon sphere monitoring} \label{sec4}
The deflection of light in an intense gravitational field exhibits two primary behaviors. When disturbances cause photons to either escape or be drawn into a black hole, the photon sphere is deemed unstable. Conversely, in cases where light remains confined and cannot escape, the photon sphere is categorized as stable, often leading to a destabilization of spacetime itself. The unstable photon sphere is essential for analyzing black hole shadows, whereas the stable photon sphere highlights regions prone to instability.
Conventional studies on photon spheres typically involve deriving the Lagrangian from the action and constructing the Hamiltonian. This process then facilitates the formulation of an effective potential, which depends on both the energy and angular momentum of the particles. The photon sphere is analyzed based on these parameters.
In contrast, this study employs a different approach, using an alternative effective potential for ultra-compact objects with spherical symmetry \cite{ph1,ph2}. This methodology was later applied in investigations of photon spheres associated with four-dimensional black holes in various spacetimes, including AdS and dS settings \cite{ph3}. A key advantage of this refined potential lies in its dependence solely on the spacetime geometry, bypassing the reliance on particle-specific parameters like energy and angular momentum. Additionally, employing a mapping technique using the equatorial plane for the vector field \(\phi\) reduces dimensional complexity, allowing for a nuanced classification of spacetime regions around compact objects \cite{ph1,ph2,ph3,ph4,ph5}.
The formulation begins by expressing the vector field \(\phi\) in terms of two distinct components, \(\phi^r\) and \(\phi^\theta\),
\begin{equation}\label{eq12}
\phi = (\phi^r, \phi^\theta).
\end{equation}
It is further expressed as \(\phi = ||\phi|| e^{i\Theta}\), where \(||\phi||\) is the magnitude of the vector. The normalized vector, denoted as \(n^a\), is defined as,
\begin{equation}\label{eq13}
n^a = \frac{\phi^a}{||\phi||},
\end{equation}
where \(a = 1, 2\). Utilizing Noether's theorem, the conservation law for charge is given by,
\begin{equation}\label{eq14}
\partial_\nu j^\nu = 0,
\end{equation}
and the total charge is expressed as,
\begin{equation}\label{eq15}
Q = \int_{\Omega} j^0 \, d^2x,
\end{equation}
where \(j^0\) signifies the charge density. Through further derivation, the current assumes the form,
\begin{equation}\label{eq16}
j^\mu = J^\mu(X) \delta^2(\phi),
\end{equation}
with \(\delta\) representing the Dirac delta function. Consequently, the topological charge becomes,
\begin{equation}\label{eq17}
Q = \int_{\Omega} J^0(X) \delta^2(\phi) \, d^2x.
\end{equation}
Considering spherical symmetry, and the metric function in four dimensions and also based on this geometry \cite{ph1,ph2,ph3,ph4,ph5}, the effective potential eith respect to Eq. (\ref{eq4}) takes the form,
\begin{equation}\label{eq19}
H(r, \theta) =\sqrt{\frac{-g_{tt}}{g_{\varphi\varphi}}} =\frac{1}{\sin\theta} \sqrt{f(r)}.
\end{equation}
The vector field components are then determined as,
\begin{equation}\label{eq20}
\phi^r =  \sqrt{f(r)} \partial_r H(r, \theta), \quad \phi^\theta = \frac{1}{r} \partial_\theta H(r, \theta)
\end{equation}
Photon spheres are located at the critical points of the effective potential, where the vector field vanishes. At these locations, a corresponding topological charge \(Q\) can be defined. Is demonstrated that the system's behavior can be classified by analyzing winding numbers and computing the total topological charge (TTC). A TTC value of \(-1\) characterizes black holes, while a value of \(0\) signifies the presence of a naked singularity \cite{ph3}. Further studies have shown that systems with asymptotically AdS or flat spacetime geometries often correspond to TTC values of \(-1\), indicating the presence of black holes. Such systems exhibit a pronounced maximum of the effective potential beyond the event horizon, associated with instability at the photon sphere. Conversely, configurations with TTC values of \(0\) or \(1\) lack an event horizon and display naked singularities, marked by both a minimum and maximum in the effective potential, suggesting a stable photon sphere. The situation in dS spacetimes is somewhat distinct due to the presence of a cosmological horizon. This horizon typically limits the analysis to the region between the Cauchy horizon and the cosmological horizon. Observations indicate that black hole behavior is more prevalent in dS models, whereas naked singularities are less commonly observed within this framework.
Based on the theoretical concepts discussed earlier and incorporating Eqs. (\ref{eq5}), (\ref{eq19}), and (\ref{eq20}), we have obtained the subsequent results,
\begin{equation}\label{eq21}
\phi^r=\frac{\csc (\theta ) \left(5 r^5 (3 M-r)+Q^4 \mu -10 Q^2 r^4\right)}{5 r^8}
\end{equation}
and,
\begin{equation}\label{eq22}
\phi^{\theta }=-\frac{\cot (\theta ) \csc (\theta ) \sqrt{\frac{r^2}{\ell^2}-\frac{2 M}{r}-\frac{Q^4 \mu }{20 r^6}+\frac{Q^2}{r^2}+1}}{r^2}.
\end{equation}
\begin{figure}[h!]
 \begin{center}
 \subfigure[]{
 \includegraphics[height=4cm,width=5cm]{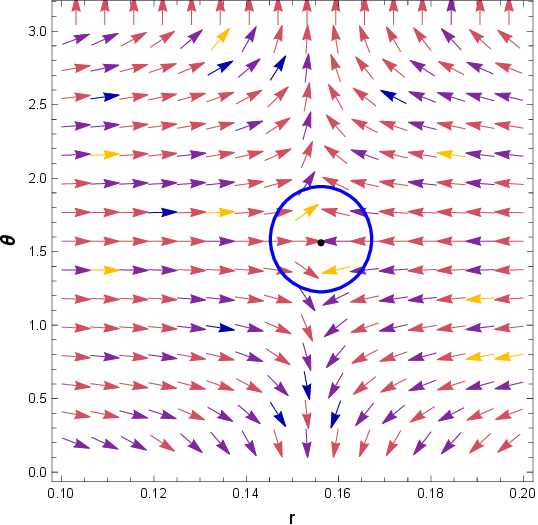}
 \label{fig4a}}
 \subfigure[]{
 \includegraphics[height=4cm,width=5cm]{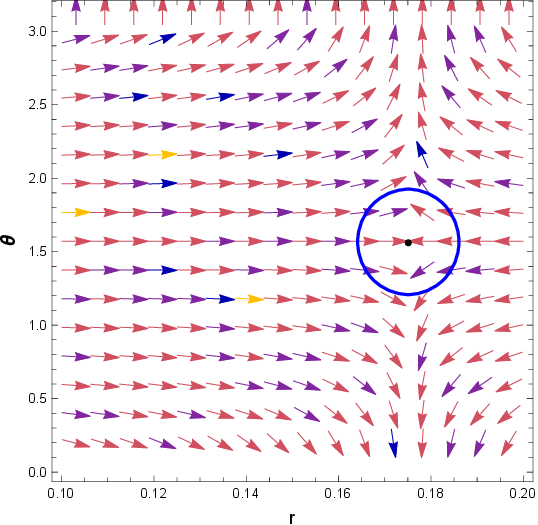}
 \label{fig4b}}
  \subfigure[]{
 \includegraphics[height=4cm,width=5cm]{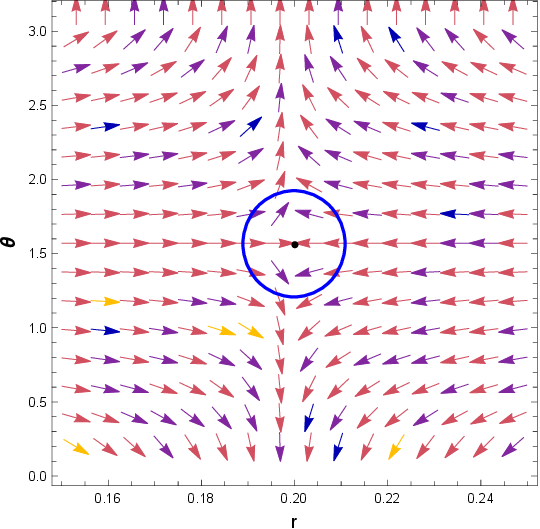}
 \label{fig4c}}
 \caption{\small{ The normal vector in the \( (r, \theta) \) plane associated with the photon spheres.(a) $Q_e=0.08$, and $M_e=0.07904$. (b) $Q_e=0.09$, and $M_e=0.08933$.(c) $Q_e=0.1$, and $M_e=0.09958$ with respect to $\ell=1$,  and $\mu=0.003$ }}
 \label{fig4}
 \end{center}
 \end{figure}

\begin{figure}[h!]
 \begin{center}
 \subfigure[]{
 \includegraphics[height=4cm,width=5cm]{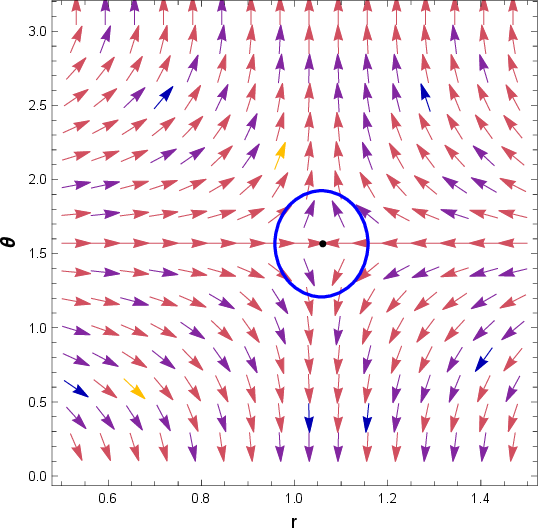}
 \label{fig5a}}
 \subfigure[]{
 \includegraphics[height=4cm,width=5cm]{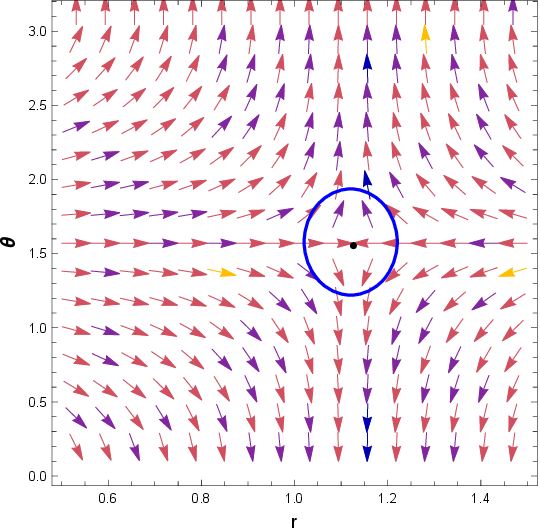}
 \label{fig5b}}
  \subfigure[]{
 \includegraphics[height=4cm,width=5cm]{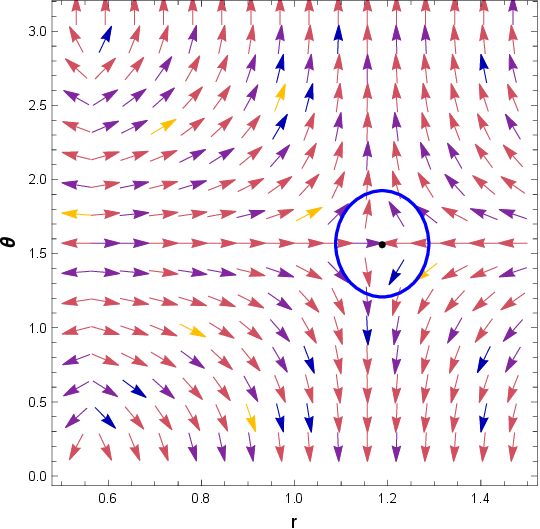}
 \label{fig5c}}
 \caption{\small{The normal vector in the \( (r, \theta) \) plane associated with the photon spheres. (a)  $Q_e=0.55$,  and $M_e=0.54272$. (b) $Q_e=0.58$,  and $M_e=0.57371$.  (c) $Q_e=0.6$,  and $M_e=0.59431$ with respect to $\ell=7$,  and $\mu=0.15$ .}}
 \label{fig5}
 \end{center}
 \end{figure}
We propose several steps to examine the compatibility of the WGC and the WCCC in the context of the specified black hole. First, it is necessary to determine the conditions under which the black hole forms event horizons. For Einstein-Euler-Heisenberg-AdS black holes, the existence of event horizons is governed by the following conditions: the black hole possesses two event horizons if \( Q^2/M^2 \leq 1 \); however, when \( Q^2/M^2 > 1 \), we proceed by analyzing the parameter space where both the WGC and WCCC are satisfied. This involves identifying constraints on the parameters \( Q \), \( M \), \( \mu \), and others to ensure the existence of event horizons and adherence to WCCC, while simultaneously ensuring that the charge-to-mass ratio complies with the WGC criteria. Such scenarios can be effectively determined by examining the black hole's metric. Building on these considerations, our goal is to explore the compatibility of the WGC with the photon spheres of Einstein-Euler-Heisenberg-AdS black holes while addressing the interplay with the WCCC across various free parameter values. Alongside detailed analyses, we will present the results concisely in Table (\ref{P3}). It is well-known that under standard conditions where \( M > Q \), the black hole model typically exhibits the expected properties, such as a total charge of \( PS = -1 \). Expanding on these findings, we will evaluate the behavior of the structure in scenarios where \( Q_e > M_e \), ensuring that the \( Q > M \) configuration satisfies both the WCCC and photon sphere conditions. In such cases, the black hole retains its defining characteristics, namely the presence of an event horizon and a photon sphere with a total negative charge of \( -1 \). To further validate this, we will analyze various structures and their behavior based on the values of free parameters, as illustrated in the figures and table. As shown in Fig. (\ref{fig4}) and Fig. (\ref{fig5}), when \( \mu \) approaches small values, the total charge of the photon sphere equals \( -1 \). This indicates that, even in the extremal state of the black hole, and when \( q/m > (Q/M)_{e} \), the structure adheres to the WCCC and qualifies as a black hole with a photon sphere \( (PS = -1) \). In this scenario, the photon sphere of the black hole provides evidence supporting the validity of the WGC within the specified context.
\begin{table}[h!]
\centering
\setlength{\arrayrulewidth}{0.4mm} 
\setlength{\tabcolsep}{2.8pt} 
\arrayrulecolor[HTML]{000000} 
\begin{tabular}{|>{\centering\arraybackslash}m{2cm}|>{\centering\arraybackslash}m{2cm}|>{\centering\arraybackslash}m{2cm}|>{\centering\arraybackslash}m{2cm}|>{\centering\arraybackslash}m{2cm}|>{\centering\arraybackslash}m{2cm}|>{\centering\arraybackslash}m{2cm}|>{\centering\arraybackslash}m{2cm}|}
\hline
\rowcolor[HTML]{9FC5E8} 
\textbf{$\ell$} & \textbf{$\mu$} & \textbf{$PS$} & \textbf{$q/m>(Q/M)_{e}$} & \textbf{$PS-WGC$}\\ \hline
\rowcolor[HTML]{EAF4FC} 
1 & 0.003 & -1 & 1.0121 & $\checkmark$ \\ \hline
1 & 0.003 & -1 & 1.00748 & $\checkmark$ \\ \hline
\rowcolor[HTML]{EAF4FC}
1 & 0.003 & -1 & 1.00417054 & $\checkmark$ \\ \hline
\rowcolor[HTML]{EAF4FC}
7 & 0.15 & -1 & 1.0134 & $\checkmark$ \\ \hline
7 & 0.15 & -1 & 1.0109  & $\checkmark$ \\ \hline
\rowcolor[HTML]{EAF4FC}
7 & 0.15 & -1 & 1.00957 & $\checkmark$ \\ \hline
\rowcolor[HTML]{EAF4FC}
\end{tabular}
\caption{The condition for consistency PS-WGC}
\label{P3}
\end{table}
\section{Conclusion} \label{sec5}

In this study, we successfully addressed the fundamental challenge of reconciling the WGC and WCCC within the framework of Einstein-Euler-Heisenberg-AdS black holes in four-dimensional spacetime. By carefully analyzing the parameter space where both conjectures could coexist, we established specific conditions under which these seemingly contradictory principles can be harmonized.

Our investigation revealed that the electromagnetic self-interaction parameter $\mu$ plays a decisive role in achieving compatibility between the WGC and WCCC. We demonstrated through detailed numerical analysis that for sufficiently small values of $\mu$ relative to the AdS radius $\ell$ (specifically when $\mu \ll \ell$), both conjectures can be simultaneously satisfied. The metric function $f(r)$ given in Eq.~(\ref{eq5}) was thoroughly examined for various parameter configurations, allowing us to identify the critical behaviors that determine horizon formation and stability. As illustrated in Fig.~(\ref{m1}), the metric function exhibits distinct characteristics depending on the values of $M$, $Q$, $\mu$, and $\ell$. We systematically explored these dependencies to establish the boundary conditions where the WGC and WCCC remain consistent with each other.

Our analysis of the $(Q_e-M_e)$ plane, depicted in Figs.~(\ref{m2}) and (\ref{m3}), clearly demonstrated how the relationship between these parameters evolves with varying $\mu$ values. Notably, Fig.~(\ref{fig2b}) provided a magnified view of the critical region where compatibility between the two conjectures emerges. The numerical results, comprehensively presented in Tables (\ref{P1}) and (\ref{P2}), confirmed that for small values of $\mu$ (e.g., $\mu = 0.003$), the condition $Q_e > M_e$ can be satisfied while maintaining the structural integrity of the black hole, thereby validating both the WGC and WCCC simultaneously. e extended our investigation to examine the interplay between photon spheres and the WGC, developing a novel approach to classify these interactions. The vector field components $\phi^r$ and $\phi^\theta$, derived in Eqs.~(\ref{eq21}) and (\ref{eq22}), enabled us to characterize the topological properties of photon spheres in Einstein-Euler-Heisenberg-AdS black holes. The visualization of these properties in Figs.~(\ref{fig4}) and (\ref{fig5}) revealed that when $\mu$ approaches small values, the photon sphere maintains a total charge of $PS = -1$, confirming that even in extremal states where $q/m > (Q/M)_e$, the black hole structure preserves its essential characteristics while supporting both conjectures. Table (\ref{P3}) consolidated our findings regarding the compatibility of photon spheres with the WGC, showing that for appropriate parameter selections, both the WGC and WCCC can be satisfied while maintaining a well-defined photon sphere structure with $PS = -1$. This result is particularly significant as it provides a concrete physical interpretation of how the abstract principles of the WGC manifest in observable astrophysical phenomena. The mathematical formalism we developed, particularly the derivation of $r_e$ in Eq.~(\ref{eq71}) and $M_e$ in Eq.~(\ref{eq72}), established a rigorous foundation for determining the extremality conditions where both conjectures can coexist. These equations, though complex, allowed us to precisely identify the parameter regions where the simultaneous validity of the WGC and WCCC occurs.

Our most significant finding was that the Einstein-Euler-Heisenberg-AdS black hole model offers a natural framework for resolving the tension between the WGC and WCCC. Unlike many other black hole models where these conjectures remain fundamentally incompatible, the nonlinear electrodynamic effects captured by the Euler-Heisenberg term provide the necessary physical mechanism to reconcile these principles. The critical condition $q^2/m^2 \geq (Q^2/M^2)_e$ emerged as the key criterion determining compatibility. The validation of both conjectures within this model has profound implications for quantum gravity theories, suggesting that nonlinear electromagnetic effects may play a crucial role in the transition between classical and quantum gravitational regimes. Our study demonstrated that the Einstein-Euler-Heisenberg-AdS black hole represents a promising candidate for further exploring the connections between these regimes, potentially offering insights into how quantum gravity principles manifest in macroscopic systems. Furthermore, we established that the photon sphere characteristics of these black holes provide an additional observational window into the validity of the WGC. The consistent behavior of photon spheres across parameter spaces where both conjectures hold suggests that astrophysical observations of black hole shadows and related phenomena may eventually offer empirical tests of these theoretical principles.

Looking forward, this study opens several promising avenues for future research. A natural extension would be to investigate whether similar compatibility patterns emerge in other modified gravity theories or alternative nonlinear electrodynamic models. The methodology we developed for analyzing the interplay between the WGC, WCCC, and photon spheres could be applied to a broader class of black hole solutions, potentially establishing a comprehensive classification system based on their compatibility with swampland conjectures. Future work could also explore the thermodynamic implications of the parameter regions where both conjectures hold. Understanding how Hawking radiation, black hole evaporation, and information paradox considerations interact with the WGC-WCCC compatibility could yield deeper insights into quantum information aspects of black hole physics. Additionally, extending our analysis to rotating black holes and more complex spacetime geometries would provide a more complete picture of how these conjectures operate in realistic astrophysical scenarios. Finally, our findings suggest that observational astronomy could potentially contribute to testing these theoretical principles. Developing specific observational signatures that could distinguish between black hole models where the WGC and WCCC are compatible versus those where they remain in tension would represent a significant advancement in connecting abstract theoretical physics with empirical science. The explicit connection we established between photon sphere properties and the validity of these conjectures represents a first step in this direction, potentially leading to observational tests of quantum gravity principles through precision black hole imaging techniques.

\section{Appendix}
Using equations (\ref{eq1}), (\ref{eq2}), and (\ref{eq3}), we calculate the value of \( r_{\text{e}} \) in the following form
\begin{equation}\label{eq71}
\begin{split}
&r_e=\bigg\{-\frac{l^2}{12}+\frac{1}{2} \bigg[\frac{l^4}{36}+\frac{2 l^2 Q^2}{9}+\\&\frac{\bigg(-128 l^6 Q^6+432 l^6 Q^4 \mu +3456 l^4 Q^6 \mu +\big(\left(-128 l^6 Q^6+432 l^6 Q^4 \mu +3456 l^4 Q^6 \mu \right)^2-4 \left(16 l^4 Q^4+144 l^2 Q^4 \mu \right)^3\big)^{\frac{1}{2}}\bigg)^{\frac{1}{3}}}{36 \sqrt[3]{2}}+\\&\frac{4 \sqrt[3]{2} \left(l^4 Q^4+9 l^2 Q^4 \mu \right)}{9 \bigg(-128 l^6 Q^6+432 l^6 Q^4 \mu +3456 l^4 Q^6 \mu +\big(\left(-128 l^6 Q^6+432 l^6 Q^4 \mu +3456 l^4 Q^6 \mu \right)^2-4 \left(16 l^4 Q^4+144 l^2 Q^4 \mu \right)^3\big)^{\frac{1}{2}}\bigg)^{\frac{1}{3}}}\bigg]^{\frac{1}{2}}+\\&\frac{1}{2} \bigg\{\frac{l^4}{18}+\frac{4 l^2 Q^2}{9}-\\&\frac{\bigg(-128 l^6 Q^6+432 l^6 Q^4 \mu +3456 l^4 Q^6 \mu +\big(\left(-128 l^6 Q^6+432 l^6 Q^4 \mu +3456 l^4 Q^6 \mu \right)^2-4 \left(16 l^4 Q^4+144 l^2 Q^4 \mu \right)^3\big)^{\frac{1}{2}}\bigg)^{\frac{1}{3}}}{36 \sqrt[3]{2}}-\\&\frac{4 \sqrt[3]{2} \left(l^4 Q^4+9 l^2 Q^4 \mu \right)}{9 \bigg(-128 l^6 Q^6+432 l^6 Q^4 \mu +3456 l^4 Q^6 \mu +\big(\left(-128 l^6 Q^6+432 l^6 Q^4 \mu +3456 l^4 Q^6 \mu \right)^2-4 \left(16 l^4 Q^4+144 l^2 Q^4 \mu \right)^3\big)^{\frac{1}{2}}\bigg)^{\frac{1}{3}}}+\\&\frac{l^6}{27}-\frac{4 l^4 Q^2}{9}\bigg/\bigg[4 \bigg(\frac{l^4}{36}+\frac{2 l^2 Q^2}{9}+\\&\frac{\bigg(-128 l^6 Q^6+432 l^6 Q^4 \mu +3456 l^4 Q^6 \mu +\big(\left(-128 l^6 Q^6+432 l^6 Q^4 \mu +3456 l^4 Q^6 \mu \right)^2-4 \left(16 l^4 Q^4+144 l^2 Q^4 \mu \right)^3\big)^{\frac{1}{2}}\bigg)^{\frac{1}{3}}}{36 \sqrt[3]{2}}+\\&\frac{4 \sqrt[3]{2} \left(l^4 Q^4+9 l^2 Q^4 \mu \right)}{9 \bigg(-128 l^6 Q^6+432 l^6 Q^4 \mu +3456 l^4 Q^6 \mu +\big(\left(-128 l^6 Q^6+432 l^6 Q^4 \mu +3456 l^4 Q^6 \mu \right)^2-4 \left(16 l^4 Q^4+144 l^2 Q^4 \mu \right)^3\big)^{\frac{1}{2}}\bigg)^{\frac{1}{3}}}\bigg)^{\frac{1}{2}}\bigg]\bigg\}^{\frac{1}{2}}\bigg\}^{\frac{1}{2}}
\end{split}
\end{equation}
Also, we can calculate the $M_{e}$ as follows,
\begin{equation}\label{eq72}
\begin{split}
M_e=\frac{r_e \left(r_e^2+l^2\right)}{2 l^2}-\frac{\mu  Q_e^4}{40 r_e^5}+\frac{Q_e^2}{2 r_e}
\end{split}
\end{equation}

\end{document}